\documentclass[a4paper,fleqn]{cas-sc}

\usepackage[authoryear]{natbib}
\usepackage{commath}

\def\tsc#1{\csdef{#1}{\textsc{\lowercase{#1}}\xspace}}
\tsc{WGM}
\tsc{QE}
\tsc{EP}
\tsc{PMS}
\tsc{BEC}
\tsc{DE}

\begin{document}
\let\WriteBookmarks\relax
\def\floatpagepagefraction{1}
\def\textpagefraction{.001}
\shorttitle{ML prediction of electrostatic charge distribution}
\shortauthors{C. Wilms et~al.}

\title [mode = title]{ML enhanced measurement of the electrostatic charge distribution of powder conveyed through a duct}                      



\affiliation[1]{organization={Physikalisch-Technische Bundesanstalt (PTB)},
                addressline={Bundesallee 100}, 
                postcode={38116}, 
                postcodesep={}, 
                city={Braunschweig},
                country={Germany}}
                
\affiliation[2]{organization={Otto von Guericke University of Magdeburg, Institute of Apparatus and Environmental Technology},
                addressline={\mbox{Universitätsplatz 2}}, 
                postcode={39106}, 
                postcodesep={}, 
                city={Magdeburg},
                country={Germany}}
                
\affiliation[3]{organization={Physikalisch-Technische Bundesanstalt (PTB)},
                addressline={Abbestraße 2–12}, 
                postcode={10587},
                postcodesep={}, 
                city={Berlin},
                country={Germany}}

\author[1,2]{C. Wilms}[type=editor,
                        orcid=0009-0007-9643-9390]
\cormark[1]
\ead{christoph.wilms@ptb.de}
\ead[url]{https://www.ptb.de/cms/asep}
\credit{Conceptualization of this study, Methodology, Software, Writing (Original draft preparation, review and editing)}

\author[1]{W. Xu}[orcid=0000-0003-3105-4848]
\credit{Software, Writing (review and editing)}

\author[1,2]{G. Ozler}[orcid=0000-0002-4196-7600]
\credit{Software, Writing (review and editing}

\author[1]{S. Janta\v{c}}[orcid=0000-0001-5682-2255]
\credit{Software, Writing (review and editing)}

\author[3]{S. Schmelter}[orcid=0000-0001-8426-9602]
\credit{Supervision, Writing (review and editing)}

\author[1,2]{H. Grosshans}[orcid=0000-0001-6441-7225]
\credit{Funding acquisition, Software, Supervision, Writing (review and editing)}


\cortext[cor1]{Corresponding author}


\begin{abstract}
The electrostatic charge acquired by powders during transport through ducts can cause devastating dust explosions.
Our recently developed laser-optical measurement technique can resolve the powder charge along a one-dimensional (1D) path.
However, the charge across the duct’s complete two-dimensional (2D) cross-section, which is the critical parameter for process safety, is generally unavailable due to limited optical access.
To estimate the complete powder charge distribution in a conveying duct, we propose a machine learning (ML) approach using a shallow neural network (SNN).
The ML algorithm is trained with cross-sectional data extracted from four different three-dimensional direct numerical simulations of a turbulent duct flow with varying particle size.
Through this training with simulation data, the ML algorithm can estimate the powder charge distribution in the duct’s cross-section based on only 1D measurements.
The results reveal an average $L^1$-error of the reconstructed 2D cross-section of 1.63\,\%.
\end{abstract}


\maketitle

\section{Introduction}
Pneumatic conveying is widely applied in industry to transport material, usually powders and particulate materials, using fluids like air as a conveying medium. Such transport systems are superior to mechanical conveyors regarding automation, maintenance and installation cost, and layout flexibility \citep{Lim2006}. However, electrostatic charging can cause a severe hazard in operation as a discharge can ignite a dust explosion \citep{Ding2024}. Hence, pneumatic conveying systems need to be well-controlled to counteract this risk. Therefore, at least two issues are crucial: 
(a) knowledge about the boundaries for safe operation as a function of material properties, ambient conditions, etc., and (b) precise monitoring to assess if the system is within the safety conditions.

At the moment, the charge of particles $Q$ can be measured non-invasive using Faraday pails or by grounding the conveying duct, resulting in an electric current caused by the charge transfer when particles pass the measurement section \citep{Matsusaka2006, Ndama2011}. Even though these principles are simple and cheap, they report only the average charge of the particles, which can significantly differ from the actual charge. For example, the charge of bipolar particles would cancel out, leading to a close to zero Faraday reading. In terms of safety, however, this would lead to a fallacy since local and instantaneous charge extremes can cause a discharge.
To address this problem, a spatially resolved measurement of the particle charge distribution would be necessary. \citet{Xu2024} presented a new technology that can measure the time-averaged particle charge distribution on a one-dimensional (1D) path oriented in a wall-normal direction. To this end, multiple particles are illuminated and tracked within a two-dimensional (2D) slice of a finite thickness. In the second step, two transparent electrodes on opposite sides of the duct apply a homogeneous electric field. Charged particles respond to the applied electric field with an increased response for stronger charged particles. By subtracting the time-averaged measurements of the particle velocity and particle acceleration without an electric field, the mean forces of the fluid on the particles cancel out, allowing us to determine the electric charge of particles.

The disadvantage of the laser-based charge measurement is the effort required to quantify the charge in the complete cross-section.
The system cannot measure a cross-section in one step.
Instead, the laser plane must be offset to scan the section by several measurements.
Such an effort seems unfeasible in industrial applications, especially since it requires large optical access to the flow.

To enhance the method by \citet{Xu2024}, this paper proposes a data-driven approach that predicts the entire streamwise-normal charge distribution based on a single line measurement. 
Computational fluid dynamic (CFD) simulations of a particle-laden turbulent duct flow provided the necessary training data. 
In the following, the developed machine learning (ML) algorithm is presented, and its potential is evaluated. 
This procedure prepares the laser-based particle charge measurement for industrial application and contributes to reducing hazards when operating pneumatic conveying systems.

\section{Method}
This section describes, in \ref{sec:method_CFD}, the setup of the duct flow simulations and, in \ref{sec:method_ML} the ML algorithm for predicting the particle charge distribution and in \ref{sec:method_trainVal} the training and validation data extracted from the CFD simulations.

\subsection{CFD simulation} \label{sec:method_CFD}
The training and validation data was extracted from four direct numerical simulations of a turbulent duct flow at a bulk Reynolds number of
\begin{equation}
	 Re=\frac{h^*u_\textrm{b}^*\rho_\textrm{f}^*}{\mu^*}=2700
\end{equation}
with half duct height $h^*$, bulk velocity $u_\textrm{b}^*$, fluid density $\rho_\textrm{f}^*$, and dynamic viscosity $\mu^*$. The $^*$ marks dimensional quantities. The friction Reynolds number amounts to
\begin{equation}
	Re_\tau=\frac{h^*u_\tau^*\rho_\textrm{f}^*}{\mu^*} \approx 180
\end{equation}

with the friction velocity $u_\tau^*$. The particles have a material density of $\rho_\textrm{p} = \rho_\textrm{p}^* / \rho_\textrm{f}^*= 1150 / 1.2$, a coefficient of restitution $e=1.0$, and a number density of $N_\textrm{p} = 800 h^{-3}$ with $h = h^* / h^*$. The size of the particles was varied between $D_\textrm{p} = D_\textrm{p}^* / h^* = 2.5 {\times} 10^{-3}$, $5 {\times} 10^{-3}$,$7.5 {\times} 10^{-3}$, and $1{\times} 10^{-2}$ in the four simulations.

The simulations were performed using the finite difference open-source solver \textit{pafiX} \citep{Grosshans2017, Grosshans2020}. The motion of the gaseous phase is described by the Navier-Stokes equations and the electrostatic field by Gauss law, both in an Eulerian framework on a staggered grid. The particles' motion due to the electrostatic field is calculated using Newton’s law of motion in a Lagrangian framework. The coupling between both phases is four-way, meaning that the fluid acts on the particles, the particles on the fluid, and the particles between each other and the walls, see \citet{Grosshans2017}. The fluid actions on the particles with lift \citep{Saffman1965, Mei1992} and drag \citep{Schiller1933} forces. Further details of the solver can be found in \citet{Grosshans2017} and \citet{Grosshans2020}.\\
The resolution of the Eulerian grid amounted to $256 \times 144 \times 144$ grid points in $x^*$, $y^*$, and $z^*$ direction, corresponding to $\Delta x^+ = 8.44$, $\Delta y_\textrm{c}^+ = \Delta z_\textrm{c}^+ = 3.93$ in the duct's center, and $\Delta y_\textrm{w}^+ = \Delta z_w^+ = 0.05$ at the wall. The $^+$ indicates wall units which are achieved by normalization with $u_\tau^*$ and kinematic viscosity $\nu^* = \mu^*/\rho_\textrm{f}^*$. The coordinates are normalized with $h^*$. The computational domain had a size of $12h^*$ $\times$ $2h^*$ $\times$ $2h^*$ with cyclic boundary conditions in the streamwise direction and walls in the spanwise directions. The discretization of the governing equations is of second order in space and time. The time stepping is dynamically adjusted to a constant CFL number of 0.25.\\
At timestep zero, uncharged particles are seeded in a fully developed turbulent duct flow which has been precalculated. By collisions with the wall, the particles accumulate charge over time. Particle-particle collisions lead to a spatial distribution of the charge, but not to a change in the overall charge of the system. This process is modeled by a simple and robust approach where the charge transfer $\Delta Q^*$ from the wall to a particle is described by
\begin{equation}
\Delta Q_n^* = C_1 A_\mathrm{pw}^* \left( Q_{\mathrm{sat}}^* - Q_n^* \right)
\end{equation}
and during inter-particle contact by
\begin{equation}
\Delta Q_n^* = C_2 A_\mathrm{pp}^* \left( Q_m^* - Q_n^* \right).
\end{equation}
The charge of particles before the impact is denoted as $Q_n^*$ and $Q_m^*$, the ratio of the contact area to total particle surface area is given as $A_\mathrm{pw}^*$ and $A_\mathrm{pp}^*$ and is calculated by a Hertzian collision model. $C_1$ and $C_2$ are proportionality factors, in this case 0.1.

In total, $7{\times} 10^4$ time steps were computed. This covers the time until the first particles reach a predefined saturation charge of $Q_{\textrm{sat}}^* = 50$\,fC and the average charge is around 80\,\% of $Q_{\textrm{sat}}^*$. In the following, the particle charge $Q^*$ is displayed in normalized form as $Q = Q^*/Q_{\textrm{sat}}^*$, see figure \ref{fig:particle_to_grid}(a) and \ref{fig:particle_to_grid}(c).

\subsection{ML algorithm} \label{sec:method_ML}
This section describes the data extraction from the simulation and the ML model based on a neural network.

\begin{figure}[b]
	\centering
	\resizebox{.8\textwidth}{!}{\input{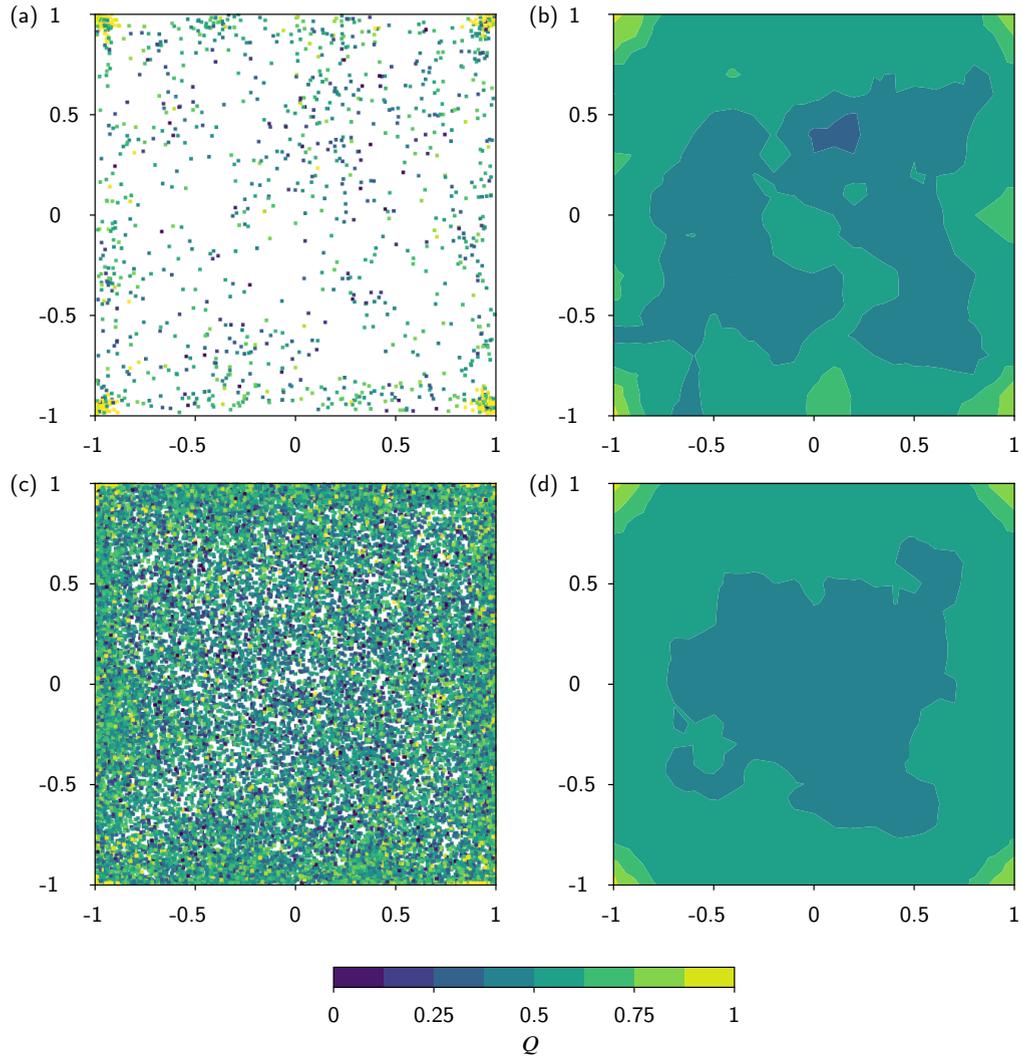}}
	\caption{Particles colored according to their charge for (a) a streamwise-normal slice with a thickness of $h=3/8$ and (c) of all particles within the duct. The mapping of the particles in (a) to a regular grid is depicted by (b), and (c) corresponds to (d).}
	\label{fig:particle_to_grid}
\end{figure}

Figure \ref{fig:particle_to_grid}(a) shows the particles with their charge for a streamwise-normal slice with a thickness of $h=3/8$ while subfigure (c) renders all particles within the duct. 
Snapshots of the Lagrangian particle field were saved every 1000 time steps of the simulation, corresponding to $\Delta t^+ \approx 0.52$. In total, this results in 70 snapshots per simulation. For each snapshot, the result fields were mapped on a regular equidistant Cartesian grid with $128 \times 21 \times 21$ grid points. The mapping was performed by determining a distance-weighted mean of the nearest 10 particles for each grid point. In the next step, 128 equidistant slices in streamwise direction were extracted. Figure \ref{fig:particle_to_grid}(b) depicts an instantaneous particle charge field mapped on a regular grid. To mimic the temporal averaging, which is necessary for the experimental measurement setup, spatial averaging in flow direction was applied, resulting in a single slice per time step, see figure \ref{fig:particle_to_grid}(d). These slices created the database for the ML. Hence, the database consisted of 280 two-dimensional slices with a resolution of $21 \times 21$ grid points.\\

The objective of the ML algorithm is to predict a 2D field based on the single path data. For this purpose, \citet{Erichson2020} developed a ML algorithm based on a shallow neural network (SNN), which is applied in the following.  Compared to a deep neural network (DNN), the advantages of a SNN are a smaller training dataset, faster training, and a smaller risk of overfitting. However, the few hidden layers restrict the SNN's ability to learn complex features,  e.g., vision and human language, from data which results in a risk of underfitting \citep{Bianchini2014}. It is, therefore, appropriate to transition from an SNN to a DNN in cases where the SNN is unable to sufficiently minimize the loss function.

The network decodes the input of the 1D path over two fully connected hidden layers to an output of the 2D field. The first hidden layer consisted of 40 neurons, followed by 45 in the second hidden layer. Both hidden layers feature a standard normalization across the batch as well as 10\,\% dropout. The algorithm was implemented in Python 3.10 \citep{Python} based on pytorch \citep{Pytorch} and can be found under \citet{Wilms2024}. For further details, the reader is referred to \citet{Erichson2020}. To improve the prediction, the neural network was operated on normalized datasets. Therefore, standard normalization based on the 1D path was applied to the 1D input and 2D output field. The normalization was grounded on the 1D path to be able to re-normalize the output data after prediction. For each trained model, bootstrapping with 20 repetitions of the training with new initial conditions of the neuron weights was applied to improve the robustness.

\subsection{Training and validation data} \label{sec:method_trainVal}
\begin{figure}[h!]
	\centering
	\resizebox{.7\textwidth}{!}{\input{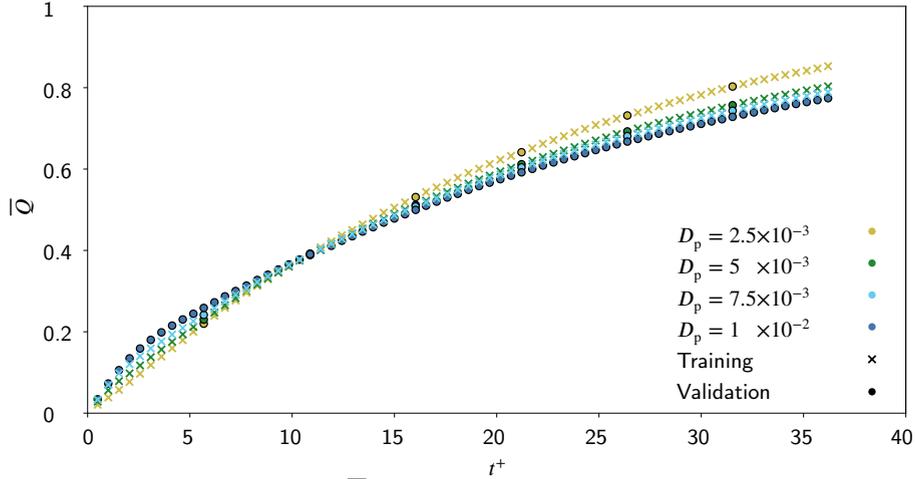}}
	\vspace{0.01cm}
	\caption{Evolution of the mean particle charge $\overline{Q}$. The marked positions (circles) represent the time steps excluded in the training and used for validation.}
	\label{fig:q_el_over_time}
\end{figure}

The training and validation data for the SNN were extracted from the four simulations described above. At the start of the simulation, uncharged particles were seeded homogeneously distributed in a fully developed turbulent duct flow. Due to triboelectricity, the average charge of all particles increased over time until it reached the predefined saturation charge of $Q_{\textrm{sat}} = 1$, see figure \ref{fig:q_el_over_time}. Particles with a smaller size reach the saturation charge faster. Three different versions of the SNN were trained to validate its performance. Therefore, the training data was varied, while the validation data was the same for all models. The validation data consists of all time steps from the case with a particle size of $D_\textrm{p} = 1{\times} 10^{-2}$ as well as specific time steps from the other three simulations, namely $t^+_{v_{1\dots6}}=5.68$, $10.85$, $16.02$, $21.19$, $26.36$, and $31.53$. This data was excluded from the training data of the three models. 

The first model $M1$ was trained on all available time steps ($0 \le t^+ \le 36.17$), the second model $M2$ was trained with a smaller training dataset containing the latest 35 time steps ($18.09 \le t^+ \le 36.17$), and the third model $M3$ was trained on the first 35 time steps ($0 \le t^+ \le 18.09$) excluding three validation time steps. Hence, $M1$ was validated on an interpolation task concerning the particle sizes $D_\textrm{p} = 2.5{\times} 10^{-3}$, $5{\times} 10^{-3}$, and $7.5{\times} 10^{-3}$ while $M2$ and $M3$ had to deal with some extrapolation in the validation and less training data. The case with a particle size of $D_\textrm{p} = 1{\times} 10^{-2}$ was an extrapolation for all three models.

For the training, from each slice, two 1D paths were extracted (at $x=0$ and $z=0$). Hence, $M1$ was trained on 384 data pairs and $M2$ and $M3$ on 192 data pairs. In the validation step, the 1D path was taken at the same positions which resulted in 176 one-dimensional validation paths (1 case with 140 paths and 3 cases with each 12 paths).

The reconstructed fields of the SNN were quantified by the $L^1$- and $\overline{Q}$-error:
\begin{equation}
	L^1\textrm{-error} = \left(\sum_{i=1}^n \abs{1 - Q_{\textrm{ref},i} / Q_{\textrm{rec},i}}\right) / n \hspace{0.5cm};\hspace{0.5cm} \overline{Q}\textrm{-error} = \abs{1 - \overline{Q_{\textrm{ref}}} / \overline{Q_{\textrm{rec}}}}
\end{equation}
where $n$ represents the number of points on the 2D grid, $Q_{\textrm{ref}}$ the ground truth 2D field and $Q_{\textrm{rec}}$ the reconstructed field. The bar $(\overline{\cdot})$ represents a spatial average. The latter metric can be compared to the state-of-the-art measurements, which supports the reliability in practical application.

\section{Results \& Discussion}
\begin{figure}[p]
	\hspace*{-0.5cm}
	\resizebox{\textwidth}{!}{\input{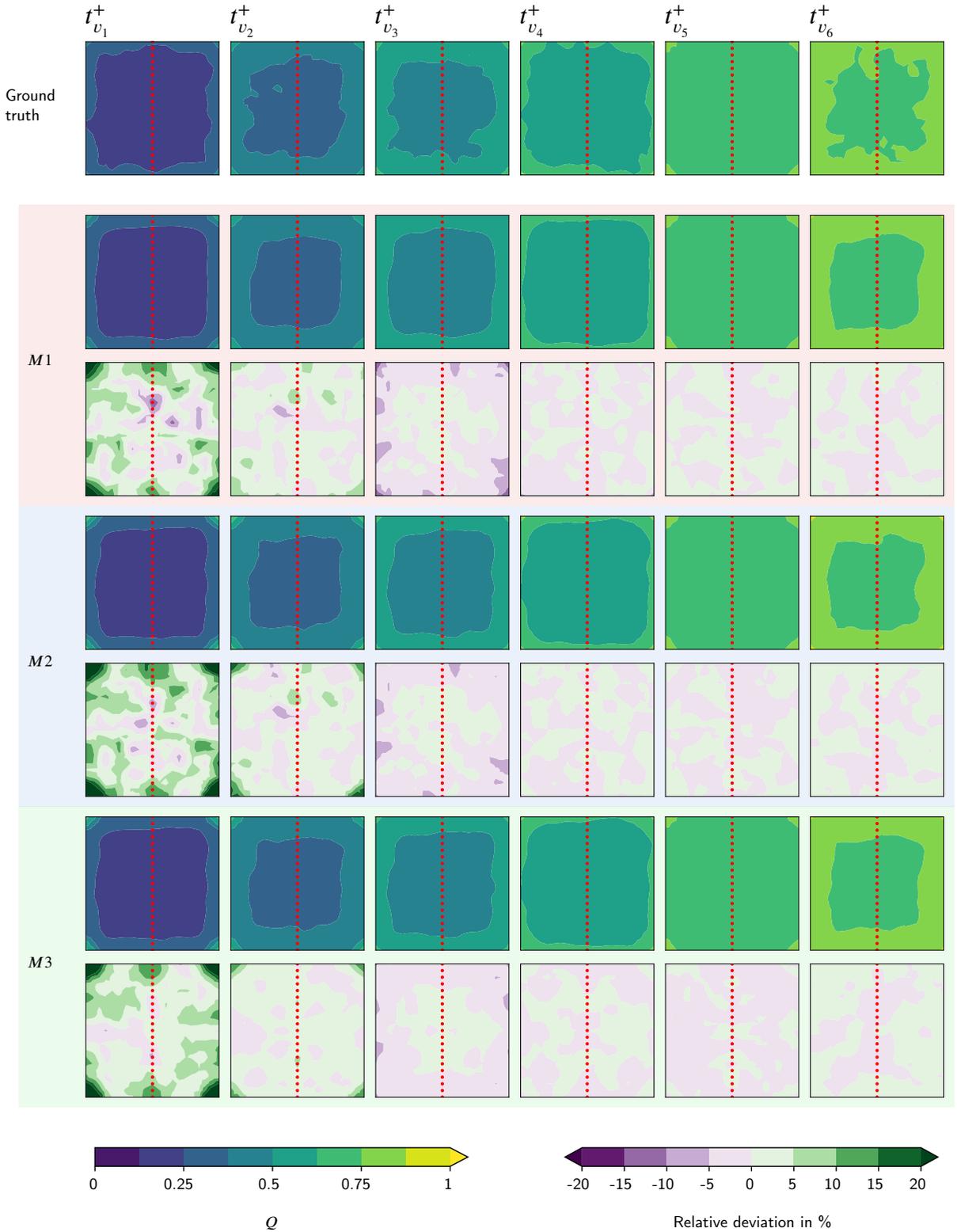}}
	\caption{Contour plots of the ground truth particle charge fields ($Q_{\textrm{ref}}$) for all validation time steps with a particle size of $D_\textrm{p} = 5{\times}10^{-3}$. The reconstructions ($Q_{\textrm{rec}}$) of the three models is shown below the first row including a contour plot of the relative deviation in the range of $\pm$\,20\,\%. The red dots indicate the location of the 1D path.}
	\label{fig:overview_reconstruction}
\end{figure}

In the first step, we analyzed the abilities of the SNN visually by comparing the reconstructed profile with the ground truth. Figure \ref{fig:overview_reconstruction} displays the results for the 1D profile extracted at the center of the duct ($y=0$) for all six validation time steps of the simulation with a particle size of $D_\textrm{p} = 5{\times}10^{-3}$ and the three models ($M1$, $M2$, $M3$). In addition, the relative error is plotted.
The analysis of the reconstructions reveals a more accurate reconstruction of the particle charge field for later time steps, independent of the model. For time steps later than $t^+_{v_2}$, every predicted point of the 2D field results in a smaller relative error than 20\,\% and for $t^+ > t^+_{v_4}$ of less than 5\,\%. The largest relative errors for the first time step occur in the corners of the cross-section. An explanation for this lies in the physical principle of particle charging and information/charge transport. The charging process starts predominantly in the corners of the duct. In such early time steps, there seems to be only a weak correlation between the initiated charging process and the charge distribution on the duct center line. Hence, it is almost impossible for the SNN to predict the corners precisely as the information is not diffused to the center.

Figure \ref{fig:L1_over_time} expands the analysis to a statistical comparison of the different models for all available validation data. Therefore, the $L^1$-error per particle size is plotted over time $t^+$. First, the trend of smaller errors for later time steps (higher particle charges) is confirmed independent of the particle size. The later time steps are easier to predict due to smaller gradients over the entire field. Thus, the field is more homogeneous and covers not several decades of charge compared to the first time steps.

The comparison between the three different models reveals that the model $M3$ performs overall best with an average $L^1$-error across all validation data of 1.63\,\%, followed by $M1$ (2.26\,\%) and $M2$ (2.44\,\%). The low error of $M3$ is astonishing as it was trained only on the first half of the time span ($0 \le t^+ \le 18.09$). Even in the second half of the time span ($18.09 \le t^+ \le 36.17$), where $M3$ was not trained at all, the error of $M3$ is the smallest ($M1$: 0.98\,\%, $M2$: 1.02\,\%, $M3$: 0.63\,\%). Thus, this period is a pure extrapolation for $M3$ (for $D_\textrm{p}=10^{-2}$ it is even an extrapolation for two parameters). Based on the training data, $M2$ should be optimized for this time span. An explanation for the good results of $M3$ might be its training on more noisy data as the charging process has just started. Additionally, this period features higher charge gradients reducing the risk of overfitting.

\begin{figure}[t]
    \resizebox{\textwidth}{!}{\input{figures/results/L1_over_time_latex}}
	\caption{$L^1$-error plotted for all validation data over time. Depicted are the results for the three trained models $M1$, $M2$, and $M3$.}
	\label{fig:L1_over_time}
\end{figure}

Averaging of the received particle charge fields allows a comparison to conventional Faraday pails, which improves the trustworthiness. It has to be noted that the results are only directly comparable when the particle distribution is homogeneous, otherwise, a weighted average or a prediction of the particle charge density would be necessary. Figure \ref{fig:mean_error_over_time} plots the deviation of $\overline{Q}$ as absolute error over time for the different validation cases. In general, the observations of the $L^1$-error are confirmed. However, due to the averaging, all errors are approximately halved. The $M1$ model achieves an average error of 1.04\,\%, $M2$ results in 1.04\,\% and $M3$ in 0.94\,\%.

\begin{figure}[t]
    \resizebox{\textwidth}{!}{\input{figures/results/mean_error_over_time_latex}}
	\caption{$\overline{Q}$-error plotted for all validation data over time. Depicted are the results for the three trained models $M1$, $M2$, and $M3$.}
	\label{fig:mean_error_over_time}
\end{figure}

\subsection{Training on a quarter of the field}
The simulations of this study model a monopolar charging process leading to a statistically symmetric particle charge field. Hence, all four corners of the duct feature the same distribution. If such knowledge is available a priori also in the practical application, it is wise to incorporate it. Therefore, this section studies the effect of averaging the particle charge distribution over all four corners and trains the neural network on the achieved quarter. The obtained prediction was replicated and transformed four times to obtain the full field.

\begin{figure}[tb]
    \resizebox{\textwidth}{!}{\input{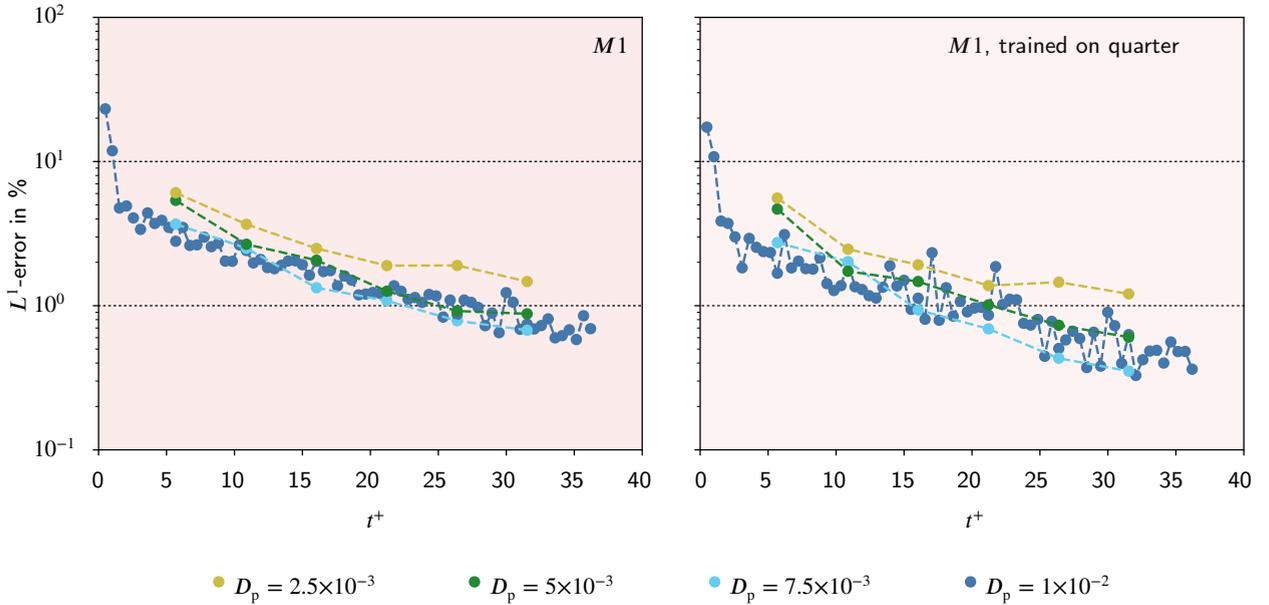}}
	\caption{$L^1$-error plotted for all validation data over time. Depicted are the results for (left) the base version of $M1$ and (right) a version of $M1$ trained on a quarter of the field.}
	\label{fig:L1_over_time_M1_vs_M1meanCorner}
\end{figure}

Figure \ref{fig:L1_over_time_M1_vs_M1meanCorner} compares the $L^1$-error of the base $M1$ model with a $M1$ model that is trained on a quarter of the field. One could observe a reduction of the mean $L^1$-error from 2.26\,\% to 1.63\,\%, which equates to a reduction of 28\,\%. This reduction seems independent of the particle size and the time step. Nonetheless, the error fluctuation across different time steps rises slightly with the latter model. As introduced before, the error reduction could be expected for two reasons: (a) the contour plots of the data are smoother due to the additional averaging step, and (b) the model is faced with a smaller decoding task. The base $M1$ model scales 21 data points from the 1D path up to 441 points on the 2D field, which corresponds to a factor of 21. The model which operates only on a quarter of the channel exhibits an upscale factor of 11 (11 points on the 1D path to 121 on the 2D field).

\subsection{Incorporating particle velocity}

Another possibility to improve the algorithm's predictions is to incorporate further information on the 1D path. The measurement principle of \citet{Xu2024} directly provides the particle velocity. Hence, an additional neural network architecture has been developed which includes as the last layer a convolution layer to combine the multiple input channels (different physical fields) to a single output channel, in this case, the particle charge.

\begin{figure}[h!]
    \resizebox{\textwidth}{!}{\input{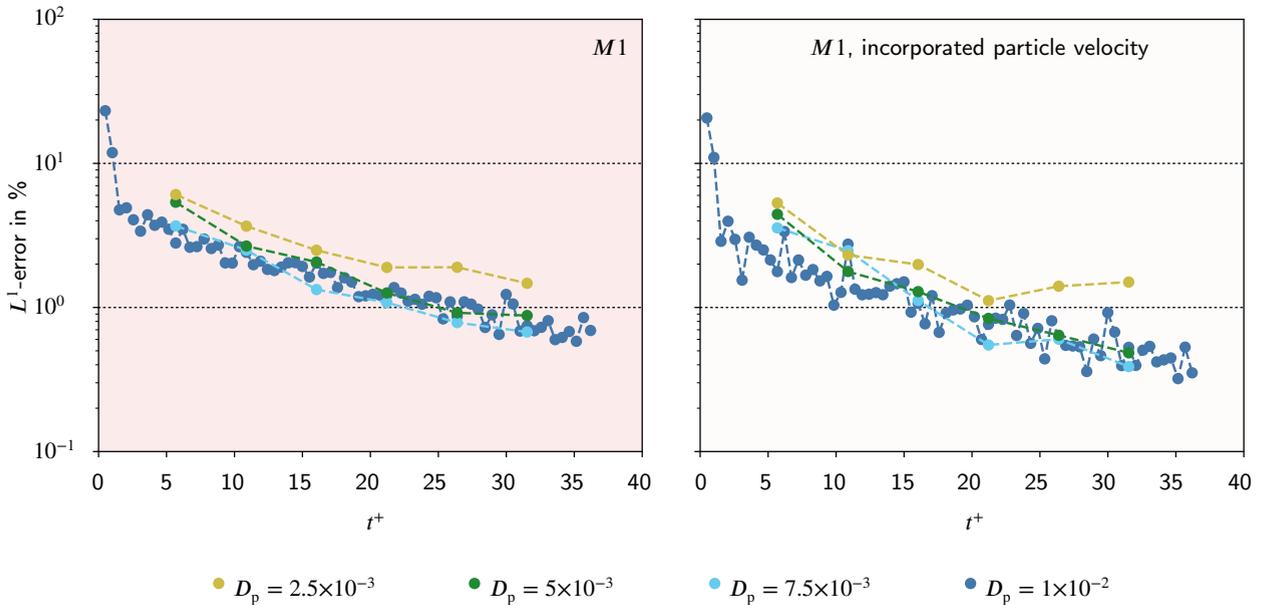}}
	\caption{$L^1$-error plotted for all validation data over time. Depicted are the results for (left) the base version of $M1$ and (right) a version of $M1$ which incorporates the particle velocity.}
	\label{fig:L1_over_time_M1_vs_M1particleVel}
\end{figure}

Figure \ref{fig:L1_over_time_M1_vs_M1particleVel} compares the $L^1$-error of the base $M1$ model with a modified version of $M1$ that includes the particle velocity. The additional information reduces the $L^1$-error by 27\,\% on average. This improvement is independent of the particle size. However, the error fluctuations increase slightly, similar to the training on a quarter of the field. The reduction of the error proves that the particle velocity gives additional information that is not directly correlated to the particle charge. This might change with the particle acceleration, which is necessary for the experiment to determine the particle charge, as it is directly correlated to the particle velocity.

\section{Conclusion \& Outlook}
This paper contributes to improving the safety of pneumatic powder conveying, an industrial transport operation. It builds upon a new measurement technology that can evaluate the particle charge distribution on a single line but not on an entire cross-section of a pipe or duct. To achieve this, we developed an algorithm to predict the 2D particle charge field based on the 1D line measurement. The algorithm's abilities were demonstrated based on simulation data, where a mean $L^1$-error of 1.63\,\% was achieved with the standard configuration. This corresponds to a mean error of 1.04\,\% of $\overline{Q}$. By incorporating the particle velocity, the $L^1$-error could be reduced additionally by 27\,\%. Considering that until now only the mean charge could be measured, this is a significant step forward in safe pneumatic conveying.
The ML model can be further improved by increasing the training database to cover a larger parameter space and increase robustness. In addition, the particle density distribution could be incorporated to calculate a weighted average charge, which would be directly comparable to a Faraday measurement.
The presented algorithm is one example of how the rise of ML can shape the future of measurement technologies impacting explosion protection.

\section*{Acknowledgements}
This project has received funding from the European Research Council (ERC) under the European Union’s Horizon 2020 research and innovation program (Grant Agreement No. 947606 PowFEct).
\printcredits

\bibliographystyle{cas-model2-names}

\bibliography{Literature.bib}

@article{Lim2006,
  title = {Effects of an electrostatic field in pneumatic conveying of granular materials through inclined and vertical pipes},
  volume = {61},
  url = {http://dx.doi.org/10.1016/j.ces.2006.07.045},
  DOI = {10.1016/j.ces.2006.07.045},
  number = {24},
  journal = {Chemical Engineering Science},
  publisher = {Elsevier BV},
  author = {Lim,  Eldin Wee Chuan and Zhang,  Yan and Wang,  Chi-Hwa},
  year = {2006},
  month = dec,
  pages = {7889–7908}
}

@article{Xu2024,
  title = {Spatially resolved measurement of the electrostatic charge of turbulent powder flows},
  volume = {65},
  ISSN = {1432-1114},
  url = {http://dx.doi.org/10.1007/s00348-024-03791-3},
  DOI = {10.1007/s00348-024-03791-3},
  number = {4},
  journal = {Experiments in Fluids},
  publisher = {Springer Science and Business Media LLC},
  author = {Xu,  Wenchao and Jantač,  Simon and Matsuyama,  Tatsushi and Grosshans,  Holger},
  year = {2024},
  month = mar 
}

@article{Grosshans2017,
  title = {Direct numerical simulation of triboelectric charging in particle-laden turbulent channel flows},
  volume = {818},
  url = {http://dx.doi.org/10.1017/jfm.2017.157},
  DOI = {10.1017/jfm.2017.157},
  journal = {Journal of Fluid Mechanics},
  publisher = {Cambridge University Press (CUP)},
  author = {Grosshans,  Holger and Papalexandris,  Miltiadis V.},
  year = {2017},
  month = apr,
  pages = {465–491}
}

@article{Erichson2020,
  doi = {10.1098/rspa.2020.0097},
  url = {https://doi.org/10.1098/rspa.2020.0097},
  year = {2020},
  month = jun,
  publisher = {The Royal Society},
  volume = {476},
  number = {2238},
  pages = {20200097},
  author = {N. Benjamin Erichson and Lionel Mathelin and Zhewei Yao and Steven L. Brunton and Michael W. Mahoney and J. Nathan Kutz},
  title = {Shallow neural networks for fluid flow reconstruction with limited sensors},
  journal = {Proceedings of the Royal Society A: Mathematical,  Physical and Engineering Sciences}
}

@article{Grosshans2020,
  title = {The effect of electrostatic charges on particle-laden duct flows},
  volume = {909},
  pages = {A21},
  url = {http://dx.doi.org/10.1017/jfm.2020.956},
  DOI = {10.1017/jfm.2020.956},
  journal = {Journal of Fluid Mechanics},
  publisher = {Cambridge University Press (CUP)},
  author = {Grosshans,  Holger and Bissinger,  Claus and Calero,  Mathieu and Papalexandris,  Miltiadis V.},
  year = {2020},
  month = dec 
}

@article{Ding2024,
  title = {Effect of airflow velocity on flame propagation and pressure of starch dust explosion in a pneumatic conveying environment},
  volume = {433},
  url = {http://dx.doi.org/10.1016/j.powtec.2023.119147},
  DOI = {10.1016/j.powtec.2023.119147},
  journal = {Powder Technology},
  publisher = {Elsevier BV},
  author = {Ding,  Jianfei and Qi,  Chang and Yan,  Xingqing and Lv,  Xianshu and Zhang,  Shuai and Liang,  He and Fan,  Tao and Yu,  Jianliang},
  year = {2024},
  month = jan,
  pages = {119147}
}

@article{Ndama2011,
  title = {A reproducible test to characterise the triboelectric charging of powders during their pneumatic transport},
  volume = {69},
  url = {http://dx.doi.org/10.1016/j.elstat.2011.03.003},
  DOI = {10.1016/j.elstat.2011.03.003},
  number = {3},
  journal = {Journal of Electrostatics},
  publisher = {Elsevier BV},
  author = {Ndama,  Adoum Traore and Guigon,  Pierre and Saleh,  Khashayar},
  year = {2011},
  month = jun,
  pages = {146–156}
}

@article{Matsusaka2006,
  title = {Simultaneous measurement of mass flow rate and charge-to-mass ratio of particles in gas–solids pipe flow},
  volume = {61},
  url = {http://dx.doi.org/10.1016/j.ces.2005.05.006},
  DOI = {10.1016/j.ces.2005.05.006},
  number = {7},
  journal = {Chemical Engineering Science},
  publisher = {Elsevier BV},
  author = {Matsusaka,  Shuji and Masuda,  Hiroaki},
  year = {2006},
  month = apr,
  pages = {2254–2261}
}

@book{Python,
 author = {Van Rossum, Guido and Drake, Fred L.},
 title = {Python 3 Reference Manual},
 year = {2009},
 isbn = {1441412697},
 publisher = {CreateSpace},
 address = {Scotts Valley, CA}
}

@incollection{Pytorch,
title = {PyTorch: An Imperative Style, High-Performance Deep Learning Library},
author = {Paszke, Adam and Gross, Sam and Massa, Francisco and Lerer, Adam and Bradbury, James and Chanan, Gregory and Killeen, Trevor and Lin, Zeming and Gimelshein, Natalia and Antiga, Luca and Desmaison, Alban and Kopf, Andreas and Yang, Edward and DeVito, Zachary and Raison, Martin and Tejani, Alykhan and Chilamkurthy, Sasank and Steiner, Benoit and Fang, Lu and Bai, Junjie and Chintala, Soumith},
booktitle = {Advances in Neural Information Processing Systems 32},
pages = {8024--8035},
year = {2019},
publisher = {Curran Associates, Inc.},
url = {http://papers.neurips.cc/paper/9015-pytorch-an-imperative-style-high-performance-deep-learning-library.pdf}
}

@article{Schiller1933,
  title={Über die grundlegenden Berechnungen bei der Schwerkraftaufbereitung},
  author={Schiller, L and Naumann, A. Z.},
  journal={Zeitung Verein Deutscher Ingenieure},
  volume={77},
  pages={318--321},
  year={1933}
}

@article{Saffman1965,
  title = {The lift on a small sphere in a slow shear flow},
  volume = {22},
  ISSN = {1469-7645},
  url = {http://dx.doi.org/10.1017/S0022112065000824},
  DOI = {10.1017/s0022112065000824},
  number = {2},
  journal = {Journal of Fluid Mechanics},
  publisher = {Cambridge University Press (CUP)},
  author = {Saffman,  P. G.},
  year = {1965},
  month = jun,
  pages = {385–400}
}

@article{Mei1992,
  title = {An approximate expression for the shear lift force on a spherical particle at finite reynolds number},
  volume = {18},
  ISSN = {0301-9322},
  url = {http://dx.doi.org/10.1016/0301-9322(92)90012-6},
  DOI = {10.1016/0301-9322(92)90012-6},
  number = {1},
  journal = {International Journal of Multiphase Flow},
  publisher = {Elsevier BV},
  author = {Mei,  R.},
  year = {1992},
  month = jan,
  pages = {145–147}
}

@article{Bianchini2014,
	title     = {On the complexity of neural network classifiers: a comparison
	between shallow and deep architectures},
	author    = {Bianchini, Monica and Scarselli, Franco},
	journal   = {IEEE Trans. Neural Netw. Learn. Syst.},
	publisher = {Institute of Electrical and Electronics Engineers (IEEE)},
	volume    =  {25},
	number    =  {8},
	pages     = {1553--1565},
	month     =  {aug},
	year      =  {2014},
	copyright = {https://ieeexplore.ieee.org/Xplorehelp/downloads/license-information/IEEE.html},
	language  = {en}
}

@software{Wilms2024,
	author       = {Wilms, Christoph and
	Xu, Wenchao and
	Oezler, Gizem and
	Jantač, Simon and
	Schmelter, Sonja and
	Grosshans, Holger},
	title        = {{ML enhanced measurement of the electrostatic 
	charge distribution of powder conveyed through a
	duct}},
	month        = oct,
	year         = 2024,
	publisher    = {Zenodo},
	version      = {v1.0.0},
	doi          = {10.5281/zenodo.14006045},
	url          = {https://doi.org/10.5281/zenodo.14006045}
}

\end{document}